\documentclass{aa}  

\usepackage{graphicx}
\usepackage[varg]{txfonts}
\usepackage[colorlinks,citecolor=blue]{hyperref}
\def\bz{$\langle$B$_z\rangle$}
\def\nz{$\langle$N$_z\rangle$}

\setcitestyle{notesep={ }}
\begin{document}

   \title{HD144941: The most extreme helium-strong star}

   \author{N. Przybilla\inst{1}
          \and 
          L. Fossati\inst{2}
          \and
          C.S. Jeffery\inst{3}
          }

   \institute{Institut f\"ur Astro- und Teilchenphysik, Universit\"at Innsbruck, Technikerstr. 25/8, 6020 Innsbruck, Austria\\
              \email{norbert.przybilla@uibk.ac.at}
         \and
             Space Research Institute, Austrian Academy of Sciences, Schmiedlstr. 6, 8042 Graz, Austria
        \and
            Armagh Observatory and Planetarium, College Hill, Armagh BT61 9DG, N. Ireland
             }

   \date{Received ; accepted }

  \abstract{Since its discovery about 50 years ago, HD144941 has generally been classified as a peculiar member of the 
  extreme helium (EHe) supergiant stars, a very rare class of low-mass hydrogen-deficient stars. We report
  the detection of a strong longitudinal magnetic field based on spectropolarimetry with FORS2 on the ESO VLT
  with surface-averaged longitudinal field strengths as large as $-$9\,kG. This is further constrained by the detection 
  of Zeeman splitting of spectral lines to a field strength of at least 15\,kG, explaining the recent finding of 
  surface spots for this star. The quantitative analysis of the stellar atmosphere based on a
  hybrid non-local thermodynamic equilibrium approach and new
  optical spectra yields an effective temperature of 22\,000$\pm$500\,K, a logarithmic surface gravity of
  4.20$\pm$0.10, and a surface helium fraction of 0.950±0.002 by number. While the metal abundances are about
  a factor of 10 sub-solar in absolute number, the metal-to-hydrogen ratios are typical of massive early-type
  stars, indicating that helium fallback in a weak, fractionated stellar wind in the presence of a magnetic
  field took place -- the canonical mechanism for the formation of the helium-strong phenomenon. Both the
  spectroscopic and the Gaia EDR3 parallax imply HD144941 to be a luminous massive star. Kinematically, we
  argue that HD144941 has reached its high Galactic latitude as a runaway star. We conclude
  that instead of being a comparatively high-gravity low-mass EHe star, HD144941 is by far the most extreme
  member of the magnetic massive helium-strong stars, with almost all atmospheric hydrogen substituted by
  helium.}


   \maketitle
%

\section{Introduction}
The helium-peculiar nature of \object{HD144941} was first noticed by \citet{MacDonnelletal70}. An early
quantitative analysis by \citet{HuKa73} confirmed it to be a member of an extreme subclass of B-type
hydrogen-deficient stars with low mass and a comparatively large radius, the so-called extreme helium (EHe)
stars \citep[see][for a review]{Jeffery08}. The first spectral analysis of HD144941 using a modern
model atmosphere code by \citet{HaJe97} and \citet{JeHa97} based on the assumption of local thermodynamic
equilibrium (LTE) established its peculiarity among the EHe stars. It
found a comparatively high gravity, an
unusually high hydrogen abundance of $\sim$5\%, a metallicity of 1.6\,dex below solar, and abundances not
showing the products of any nuclear reactions other than hydrogen burning, in contrast to the presence 
of CNO- and 3$\alpha$-process ashes in most other EHe stars. Yet,
the star appeared to show properties much closer to those of the EHe stars (with helium contents of $\sim$99\%) 
than to those of any other helium-rich stars, in particular the magnetic massive helium-strong main sequence stars, 
which show helium contents typically in the 30 to 50\% range \citep[e.g.][]{Smith96,Cidaleetal07}.
Later, non-LTE analyses confirmed the earlier findings
\citep{Przybillaetal05,Przybillaetal06,PaLa17}. The merger of two helium white dwarfs in a binary was
identified as a viable formation mechanism \citep{SaJe00}.
The recent detection of rotational modulation in a light curve of HD144941 \citep{JeRa18} measured by the
K2 mission points to the existence of surface spots, which in turn are
indicative for the presence of a magnetic field in an early-type star. In the following, we 
discuss the first spectropolarimetric observations of the star and perform a quantitative 
spectral reanalysis, which show that HD144941 is indeed a very different type of star than supposed to date.

\section{Observations}
Our target star HD144941 was observed in service mode on a total of nine nights in March, June, and July 2021 
using the FOcal Reducer and low dispersion Spectrograph 2 (FORS2) low-resolution spectropolarimeter
\citep{Appenzelleretal98} attached to the European Southern
Observatory (ESO) Very Large Telescope (VLT) Unit Telescope 1 (UT1) 
at Cerro Paranal, Chile. On the night of July 1, 2021, we obtained two sets of observations separated by 
a few hours. The data were obtained with a slit width of 0.4\arcsec\ and the grism 600B. This setting yielded a 
resolving power $R$\,$\approx$\,1700 and a wavelength coverage from 3250 to 6215\,{\AA}. On each night of 
observation, we obtained several spectra with an exposure time of 30 seconds each. We obtained a first set of 
spectra with the quarter wave plate at $-$45$^{\circ}$, then twice as many at $+$45$^{\circ}$, and then again 
exposures at $-$45$^{\circ}$, to finally obtain the same amount of spectra for each of the two angle positions. 
The observing log and Stokes $I$ signal-to-noise ratios (S/N) per pixel calculated around 4950\,\AA\ are 
listed in Table~\ref{tab:FOR2log_results}. The data were reduced using custom-made tools described in detail 
by \citet{Fossatietal15}, which are based on the recommendations of \citet{bagnulo2012}.

\begin{figure*}
\sidecaption
\centering
\includegraphics[width=11.8cm]{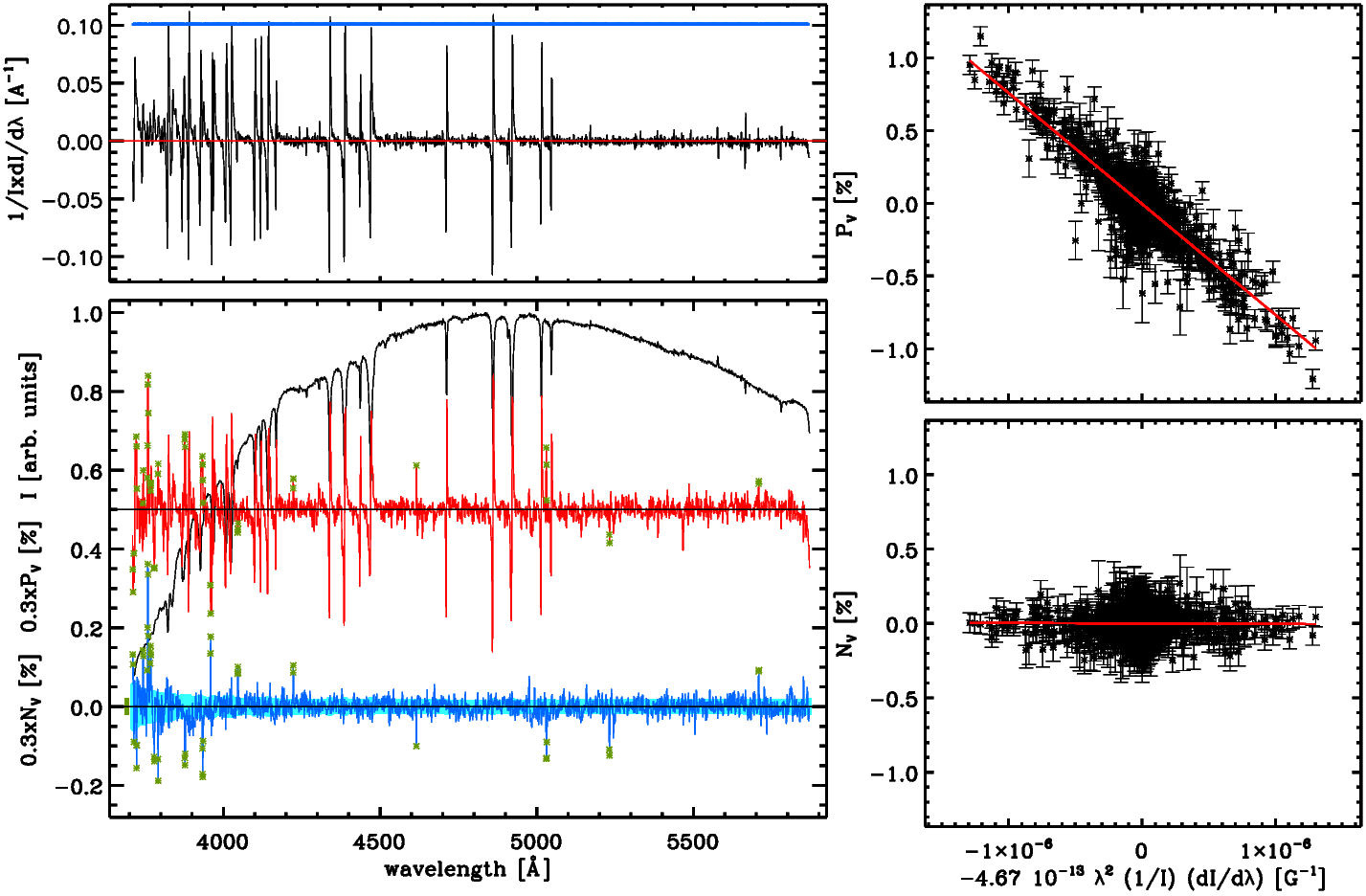}
\caption{Overview of FORS2 analysis and results from the observations of HD144941 collected on March 15, 2021
considering the full spectrum. Top left panel: Derivative of Stokes $I$. The thick blue line on top of the panel 
indicates the region used to compute \bz. Bottom left panel: From top to bottom, the panel shows the Stokes $I$ 
spectrum (black) arbitrarily normalised to the highest value, the Stokes $V$ spectrum (in \%; red) rigidly 
shifted upwards by 0.5\% for visualisation reasons, and the spectrum of the $N$ parameter (in \%; blue). The green 
asterisks mark the points that were removed by the sigma-clipping algorithm. The pale blue strip drawn on top of 
the $N$ profile shows the uncertainty associated with each spectral point. The thick green bar on the left side of 
the spectrum of the $N$ parameter shows the standard deviation of the $N$ profile. Top right panel: Linear fit from 
which \bz\ is determined. The red solid line shows the best fit giving \bz\,=\,$-$7626$\pm$86\,G. Bottom right panel: 
Same as the top right panel, but for the $N$ profile (i.e. \nz), from which we obtain \nz\,=\,$-$32$\pm$71\,G.}
\label{fig:FORS2_analysis}
\end{figure*}

\begin{table*}[ht!]
\caption{FORS2 Observing log and average longitudinal magnetic field ($\langle B_z\rangle$) values.}
\label{tab:FOR2log_results}
\centering
{\small
\begin{tabular}{lc|ccc|rr|rr}
\hline
\hline
Date & JD$-$  & No. of  & Exposure & S/N at & \bz\,(G) & \nz\,(G)     & \bz\,(G) & \nz\,(G) \\
YYYY-MM-DD     & 2450000 & frames & time (s) & 4950\,\AA & \multicolumn{2}{c|}{Hydrogen}  & \multicolumn{2}{c}{All}   \\
\hline
2021-03-15  & 9288.74493 & 40 & 30 & 1741 & $-$9004$\pm$170 & $-$129$\pm$137 & $-$7626$\pm$86  &  $-$32$\pm$71  \\ 
2021-03-24  & 9297.70603 & 40 & 30 & 1975 & $-$8146$\pm$159 &     16$\pm$135 & $-$7068$\pm$92  & $-$117$\pm$73  \\
2021-06-10  & 9376.49805 & 40 & 30 & 1307 & $-$7377$\pm$266 &    278$\pm$219 & $-$6718$\pm$134 &    154$\pm$114 \\
2021-06-12  & 9377.78924 & 40 & 30 & 1886 & $-$7963$\pm$189 &  $-$28$\pm$165 & $-$7043$\pm$85  &     97$\pm$73  \\
2021-07-01a & 9396.15904 & 40 & 30 & 2156 & $-$7549$\pm$140 &     85$\pm$117 & $-$6538$\pm$70  &     16$\pm$55  \\ 
2021-07-01b & 9396.21277 & 80 & 30 & 2761 & $-$7283$\pm$126 &  $-$48$\pm$100 & $-$6422$\pm$67  &  $-$27$\pm$51  \\
2021-07-08  & 9403.13370 & 36 & 30 & 1738 & $-$7823$\pm$164 & $-$238$\pm$116 & $-$6872$\pm$81  &  $-$80$\pm$68  \\
2021-07-11  & 9406.12985 & 36 & 30 & 1539 & $-$8375$\pm$222 & $-$799$\pm$202 & $-$6820$\pm$105 & $-$362$\pm$86  \\
2021-07-14  & 9409.14539 & 18 & 30 &  782 & $-$8382$\pm$424 &    690$\pm$304 & $-$6721$\pm$208 &    128$\pm$154 \\
2021-07-18  & 9413.10564 & 36 & 30 & 1490 & $-$8914$\pm$194 & $-$170$\pm$171 & $-$7806$\pm$103 &  $-$55$\pm$87  \\
\hline
\end{tabular}
\tablefoot{Julian date listed in the second column is that at the beginning of the exposure sequence. 
Columns three and four give the number of frames obtained during each night of observation and the exposure time of each frame, respectively. Column five lists the S/N per pixel of the intensity spectrum (Stokes $I$) calculated at $\approx$4950\,\AA\ over a wavelength range of 100\,\AA. Columns six and seven list the $\langle$B$_z\rangle$ and $\langle$N$_z\rangle$ values obtained employing the hydrogen lines from the Stokes $V$ and $N$ parameter spectra, respectively. Columns eight and nine are the same as columns six and seven, but employing the full spectrum.}} 
\end{table*}

An optical spectrum with $R$\,=\,48\,000 was obtained on April 8, 2006
using the Fibre-fed Optical Echelle Spectrograph \citep[FEROS,][]{Kauferetal99}
on the ESO-Max-Planck-Gesellschaft (MPG) 2.2m telescope at La Silla, Chile. The 3000\,s exposure yielded a peak
S/N of about 200. The data were complemented by a 780\,s exposure with
the Ultraviolet and Visual Echelle Spectrograph
\citep[UVES,][]{Dekkeretal00} on the ESO-VLT UT2 on April 10, 2006, covering the optical-UV range at
$R$\,$\approx$\,41\,000 and reaching a peak-S/N\,$\approx$\,170. The combined wavelength range at S/N high enough for a 
quality analysis covers $\sim$3200 to 8000\,{\AA}. The data reduction was performed in analogy to \citet{Kupferetal17}. 
We want to emphasise that an additional spectrum of the DC white dwarf \object{WD 1917-07} was observed with UVES to 
trace the spectral response function, facilitating the
normalisation of the target spectra in an objective way (see \citet{Koesteretal01} for details of the
method). In analogy, the spectrum of the (well-modelled) sdB star \object{HD188112} was employed for
normalisation of the FEROS data.

Low-dispersion and large-aperture spectrophotometry from the International Ultraviolet Exporer (IUE) 
was employed to constrain the spectral energy distribution (SED) of HD144941 (data sets SWP07696 and LWR06703). These were supplemented by photometry in the Johnson $UBV$-bands from
\citet{Slawsonetal92}, in the Two Micron All Sky Survey (2MASS) $JHK$-bands from \citet{2MASS2006}, and in the 
Wide-field Infrared Survey Explorer (WISE) bands from \citet{ALLWISE_vizier}.

\section{Magnetic field detection}\label{sect:bfield}
We reduced the FORS2 spectropolarimetric observations employing the pipeline described by \citet{Fossatietal15}, 
which is based on that of \citet{bagnulo2012,bagnulo2015}. Furthermore, the pipeline implements the set of tests 
recommended by \citet{bagnulo2013} to aid the identification of genuine magnetic field detections. The 
surface-averaged longitudinal magnetic field values obtained from the analysis of the FORS2 observations and 
conducted on both the Stokes $V$ spectrum and null ($N$) profile are listed in Table~\ref{tab:FOR2log_results}. 
We show in Fig.~\ref{fig:FORS2_analysis} the output of the analysis performed on the observations obtained 
on March 15, 2021, and considering the whole spectrum. The strong magnetic field detection is indicated by the 
significant slope shown in the top right panel, while the lack of a significant slope in the bottom right panel 
indicates to the first order that the magnetic field detection is genuine \citep[see][for a thorough discussion]{bagnulo2013}.

\begin{figure}
\centering
\includegraphics[width=.99\hsize]{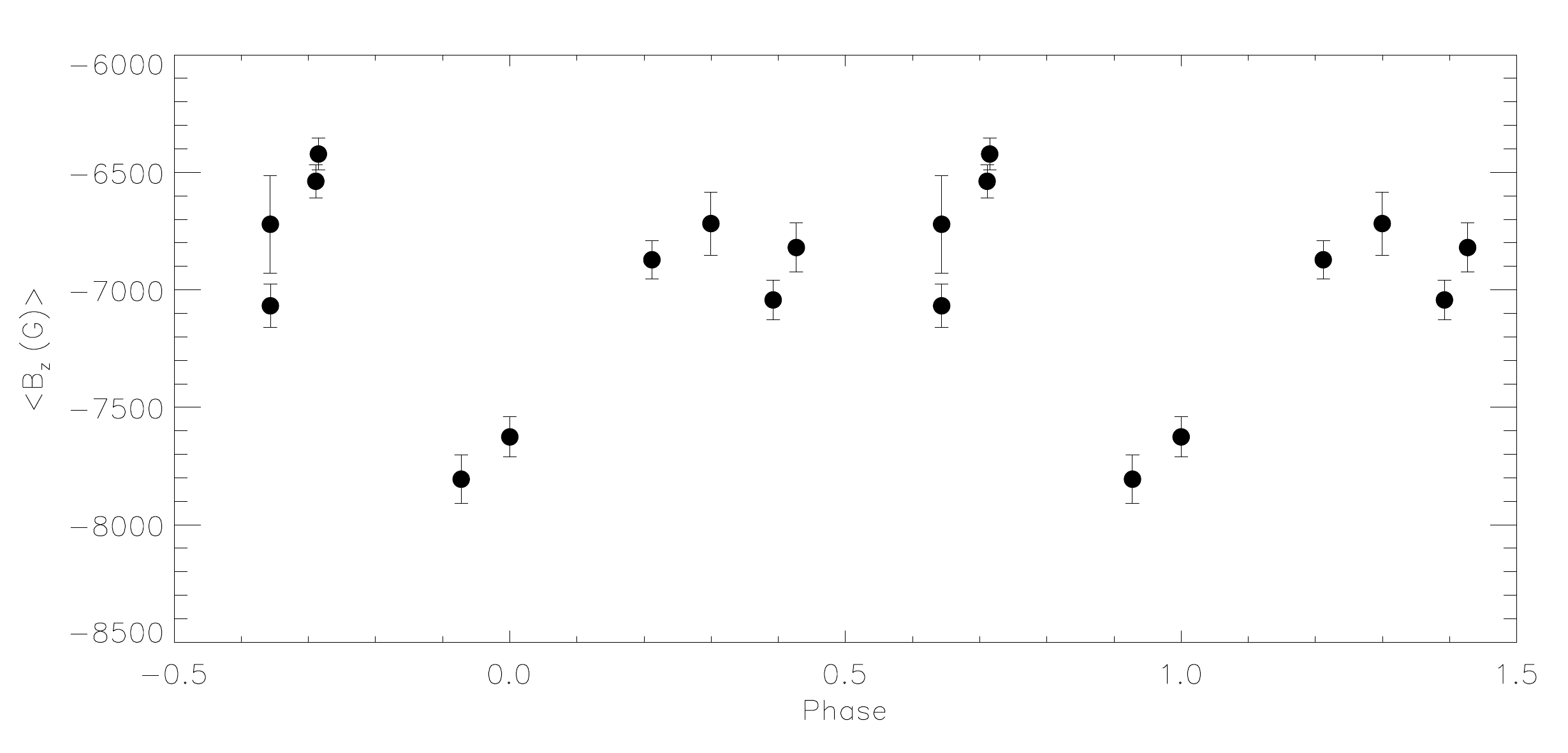}
\caption{Average longitudinal magnetic field variation of HD144941 from
measurements of the whole spectrum, phased by the 13.93\,d period derived by
\citet{JeRa18} from the K2 light curve. The zero point is given by our
first measurement, coinciding close to the maximum of the magnetic field
strength. We note the repetition of the data below phase 0 and above phase 1 for visualisation purposes.}
\label{fig:phasing}
\end{figure}

Table~\ref{tab:FOR2log_results} shows that we obtained a significant magnetic field detection 
($\gg$10$\sigma$) from each night of observation, with a peak strength of \bz\,$\approx$\,$-$9\,kG obtained 
from the data collected on March 15 and July 18, 2021, and considering the hydrogen lines. We obtained a 
significant magnetic field detection employing the hydrogen lines, or the metal lines (not shown here), or the 
whole spectrum. The \bz\ values obtained from the hydrogen lines are on average 1--1.4\,kG stronger than those 
obtained from the metal lines or the whole spectrum. For strongly magnetic stars, such discrepancies are common 
and can be ascribed, for example, to the (partial) breaking of the weak-field approximation on which the magnetic 
field measurement is based \citep{angel1970,borra1980,bagnulo2002,landstreet2014,Fossatietal15}. 

Instead, the variation of 1--2\,kG in the \bz\ values obtained from the March, June, and July data 
(see Table~\ref{tab:FOR2log_results} and Fig.~\ref{fig:phasing}) can be ascribed to rotational modulation. 
\citet{JeRa18} found a highly structured singly periodic signal in K2 observations, with a fundamental period of 
13.9$\pm$0.2\,d and multiple harmonics. We adopted the value of 13.93\,d (their Fig.~3) to phase the \bz\ values 
from analysis of the whole spectrum in Fig.~\ref{fig:phasing}. While our data are insufficient to improve on the 
period derived from the K2 light curve, they are consistent with it. As the polarity of the magnetic field does 
not change over the full period, a viewing geometry not too far from pole-on can be deduced, with the magnetic 
field also coarsely aligned with the rotation axis. We note that the structured light curve is indicative of a 
magnetic field configuration more complex than a simple dipole, but more observations are required to constrain 
the magnetic field geometry.

While the presence of a magnetic field (and rotational modulation) has not been found for any other 
EHe star, magnetic fields are crucial for the development of the He-strong phenomenon among massive main-sequence 
B-type stars. This means that the notion of HD144941 being an EHe star should be reconsidered. 

\section{Model atmosphere analysis}
For the quantitative analysis of the spectra we followed a hybrid non-LTE approach
\citep{NiPr07,Przybillaetal11} that we used before to study both EHe stars 
\citep{Kupferetal17} and helium-strong stars \citep[e.g.][]{Przybillaetal16,Castroetal17}. 
In brief, LTE model atmospheres as computed with the {\sc Atlas12} code \citep{Kurucz96} were 
used as basis for subsequent non-LTE line-formation calculations with {\sc Detail} and {\sc Surface}
\citep[][both updated and extended by K. Butler]{Giddings81,BuGi85}. The former code solves the coupled
statistical equilibrium and radiative transfer equations, while the latter computes the emergent
synthetic spectrum based on the resulting non-LTE population numbers using refined line-broadening
theories. Here, detailed broadening tables of \citet{TrBe09} were used with the model atoms for
H \citep{PrBu04}, and those of \citet{Beauchampetal97} were used for \ion{He}{i} \citep{Przybilla05}. 
We adopted model atoms for the metals according to Table~2 of \citet{Przybillaetal16}.

\begin{table}[t]
\caption[ ]{Parameters and elemental abundances of HD144941.}
\label{tab:parameters}
\centering
\setlength{\tabcolsep}{1.1mm}
{\small
\begin{tabular}{llll}
\hline
\hline
\multicolumn{4}{l}{Atmospheric parameters:}\\
$T_\mathrm{eff}$     & 22000$\pm$500\,K & $\xi$      & 2$\pm$1\,km\,s$^{-1}$\\
$\log g/\mathrm{cm\,s^{-2}}$      & 4.20$\pm$0.10    & $v \sin i$ & 7$\pm$5\,km\,s$^{-1}$\\
$y$\,(by number)     & 0.950$\pm$0.002  & $\zeta$    & 25/15$\pm$5\,km\,s$^{-1}$\tablefootmark{a}\\[1.3mm]
\multicolumn{4}{l}{Non-LTE abundances $\log n_X$, normalised to $\sum n_X$\,=\,1:}\\
H       & $-$1.30$\pm$0.02\,(4)  & Mg      & $-$5.52\,(1)\\
C       & $-$4.63$\pm$0.12\,(7)  & Al      & $-$6.64$\pm$0.06\,(4)\\
N       & $-$5.06$\pm$0.09\,(15) & Si      & $-$5.48$\pm$0.11\,(15)\\
O       & $-$4.26$\pm$0.07\,(22) & S       & $-$5.61$\pm$0.10\,(5)\\
Ne      & $-$4.67$\pm$0.04\,(2)  & Fe      & $-$5.41$\pm$0.13\,(5)\\[1.3mm]
\multicolumn{4}{l}{Photometric data:}\\
$V$      & 10.141$\pm$0.008\,mag\tablefootmark{b}    & $R_V$           & 3.8$\pm$0.1\\
$B-V$    & ~~0.034$\pm$0.007\,mag\tablefootmark{b}   & $M_V$           & $-$1.85$^{+0.10}_{-0.11}$\,mag\\
$E(B-V)$ & ~~~~0.27$\pm$0.01~~\,mag & $M_\mathrm{bol}$& $-$3.94$\pm$0.10\,mag\\[1.3mm]
\multicolumn{4}{l}{Fundamental parameters:}\\
$M/M_\odot$                & 8.1$\pm$0.3 & $\tau$                  & 11.0$^{+6.4}_{-7.8}$\,Myr\\
$R/R_\odot$                & 3.8$\pm$0.2 & $\tau/\tau_\mathrm{MS}$ & 0.29$^{+0.20}_{-0.21}$\\
$\log L/L_\odot$           & 3.47$\pm$0.04 & $P_\mathrm{rot}$      & 13.9$\pm$0.2\,d\tablefootmark{c}\\[1.3mm]
\multicolumn{4}{l}{Kinematic data:}\\
$\pi_\mathrm{EDR3}$ & ~~~0.6714$\pm$0.0282\,mas\tablefootmark{d} & $v_\mathrm{rad}$  & $-$43.7$\pm$0.5\,km\,s$^{-1}$\\
$\mu_\alpha$        & $-$6.834$\pm$0.034\,mas\,yr$^{-1}$\tablefootmark{d} & $d_\mathrm{Gaia}$ & 1556$^{+71}_{-66}$\,pc\\
$\mu_\delta$        & ~~~8.023$\pm$0.025\,mas\,yr$^{-1}$\tablefootmark{d} & $d_\mathrm{spec}$ & 1582\,$\pm$\,208\,pc\\[0.5mm]
\hline\\[-5mm]
\end{tabular}
\tablefoot{1$\sigma$-uncertainties are given. For abundances, these are
from the line-to-line scatter; systematic errors amount to an
additional $\sim$0.1\,dex.
\tablefoottext{a}{with and without considering Zeeman splitting}
\tablefoottext{b}{\citet{Slawsonetal92}}
\tablefoottext{c}{\citet{JeRa18}}
\tablefoottext{d}{\citet{Gaia16,Gaia21}}
}}
\end{table}

\begin{figure*}
\centering
\includegraphics[width=.92\hsize]{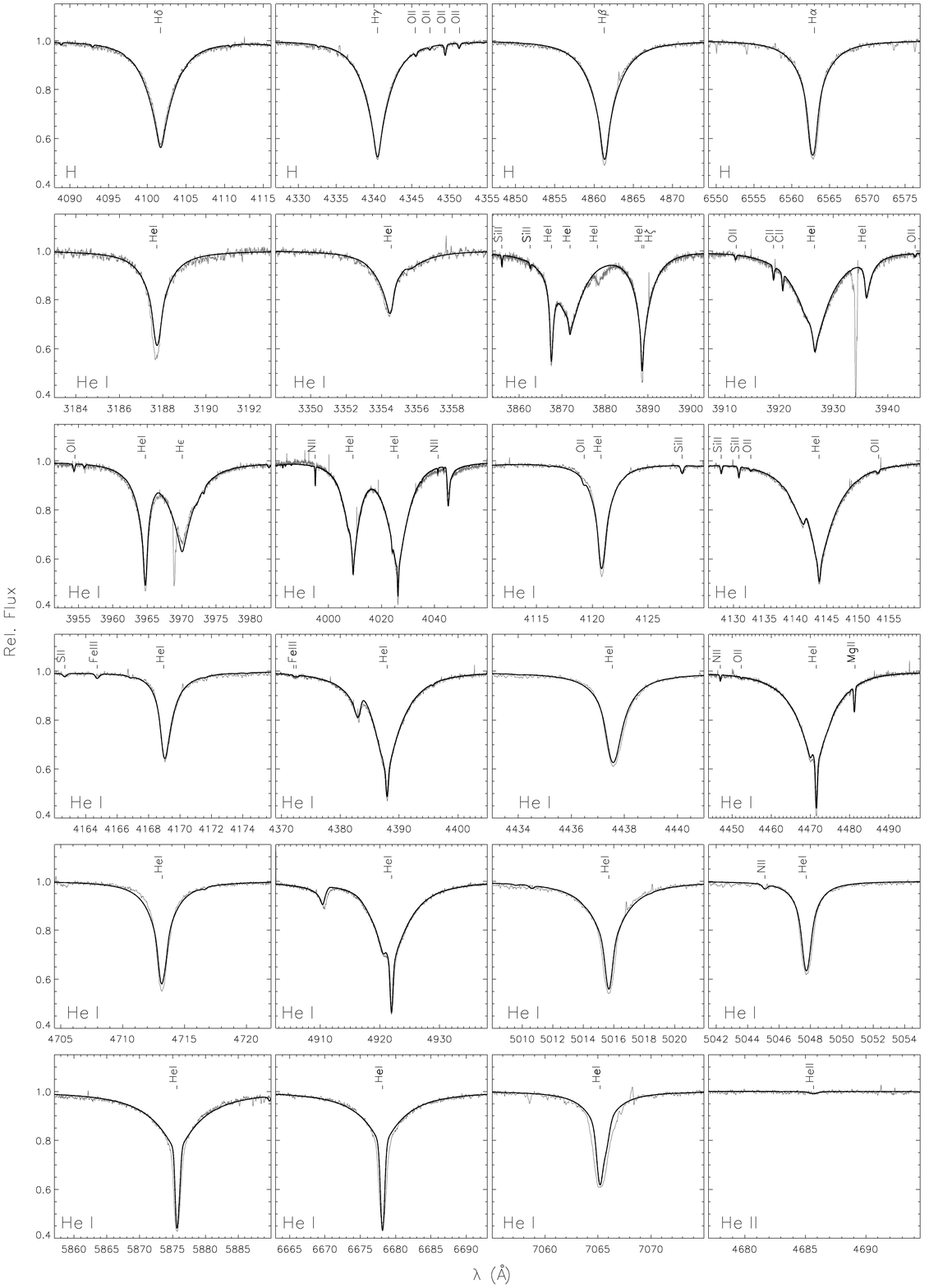}
\caption{Comparison of final model spectrum (black lines) with observed hydrogen
and helium lines in HD144941 (grey lines).}
\label{fig:spectrum_synthesis}
\end{figure*}

The atmospheric parameters, which are effective temperature $T_\mathrm{eff}$, surface
gravity $g$, helium abundance (by number) $y$,
microturbulence $\xi$, macroturbulence $\zeta$, projected rotational velocity 
$v \sin i$ as well as
abundances $\log n_X$ (number fractions, normalised to $\sum n_X$\,=\,1) of chemical
species $X$ were determined from line-profile fits. Important diagnostics for the
analysis were the Stark-broadened hydrogen Balmer and \ion{He}{i} lines, and
the ionisation balances of \ion{Si}{ii/iii} and \ion{S}{ii/iii} were established
simultaneously (i.e. abundances from different ionisation stages were brought to match). 
Independence of the abundances from the strength of the analysed lines was used as criterion to 
fix $\xi$. For a detailed description of the iterative procedure, we
invite the reader to consult \citet{NiPr12}. 

\begin{figure}[th]
\centering
\includegraphics[width=.93\hsize]{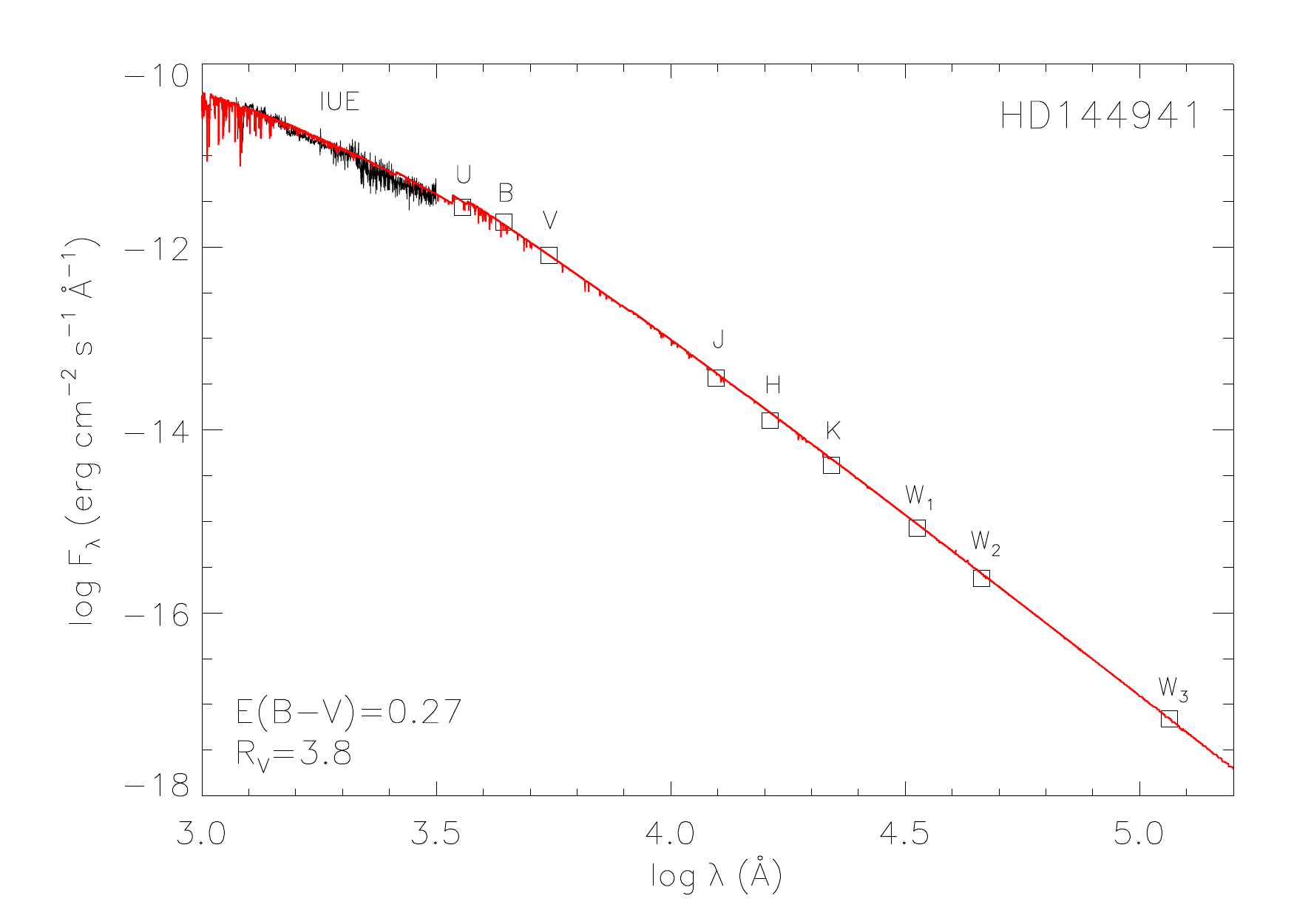}
\caption{Comparison of de-reddened spectral energy distribution of HD144941 (black line, squares) with
the model non-LTE flux (red).}
\label{fig:SED}
\end{figure}

The derived atmospheric parameters and elemental abundances are summarised in Table~\ref{tab:parameters}. 
The quality of the atmospheric parameter solution can be assessed best by the overall excellent match
between the observed and the final model spectrum. This is demonstrated in Fig.~\ref{fig:spectrum_synthesis}
for all hydrogen Balmer and \ion{He}{i} lines between 3900 and 7100\,{\AA}, plus two lines in the
optical-UV, \ion{He}{i} $\lambda\lambda$3187.74\,{\AA} and 3354.56\,{\AA}. 
\ion{He}{ii} $\lambda$4686\,{\AA} also imposes a tight upper limit on $T_\mathrm{eff}$ due to
its extreme weakness in the spectrum. Small increases in $T_\mathrm{eff}$ would lead to a more
pronounced line.
Minor deviations between observations and the model become apparent only for the lines at
$\lambda$\,$\gtrsim$\,5000\,{\AA}, such that the observed lines are broader than the model, 
increasingly so with wavelength; this point becomes significant below. Additionally, an overall good match of the metal
lines is obtained,
reproducing the observed profiles nicely by the model finally adopted, establishing ionisation equilibria of
\ion{Si}{ii/iii} and \ion{S}{ii/iii}. 
Some of the stronger lines show deviations from a rotational/macroturbulent profile in their cores,
just above the noise level -- this discussion will also be resumed later.
Noteworthy in the comparison with previous work is the large discrepancy with respect to the $\log
g$\,=\,3.45$\pm$0.15 determined by \citet{PaLa17} and our much lower microturbulence (2\,km\,s$^{-1}$ vs. 
10\,km\,s$^{-1}$ in the literature). One may speculate about continuum normalisation as a
reason for the former because of the much larger widths of the H and He lines than in typical EHe stars; 
we invite the reader to compare the widths of lines in Figs.~7 and 9
of \citet{PaLa17} with our data in Fig.~\ref{fig:spectrum_synthesis}.

Constraints on $T_\mathrm{eff}$ are also obtained from the fit of the non-LTE model fluxes to the observed 
SED of the star (see Fig.~\ref{fig:SED} for the match achieved from the UV to the thermal IR). The observed fluxes 
were de-reddened using an interstellar reddening law according to 
\citet{Cardellietal89}, fitting both the colour excess $E(B-V)$ and the ratio of
total-to-selective extinction $R_V$\,=\,$A_V/E(B-V)$\footnote{A comparatively high reddening value is found for 
the star's position 17.8\degr above the Galactic plane, 
and an anomalous reddening law. Most of the effect should stem from the well-known foreground dust lanes 
in the Scorpius region.}. The resulting extinction and the Gaia distance 
provided the absolute visual magnitude $M_V$ of the star, and the application of the
computed bolometric correction from the stellar atmosphere model provided the
bolometric magnitude $M_\mathrm{bol}$. The observed $V$ and $B-V$ data for HD144941 and the 
deduced photometric data are summarised in Table~\ref{tab:parameters}.

\begin{figure}[t]
\centering
\includegraphics[width=.98\hsize]{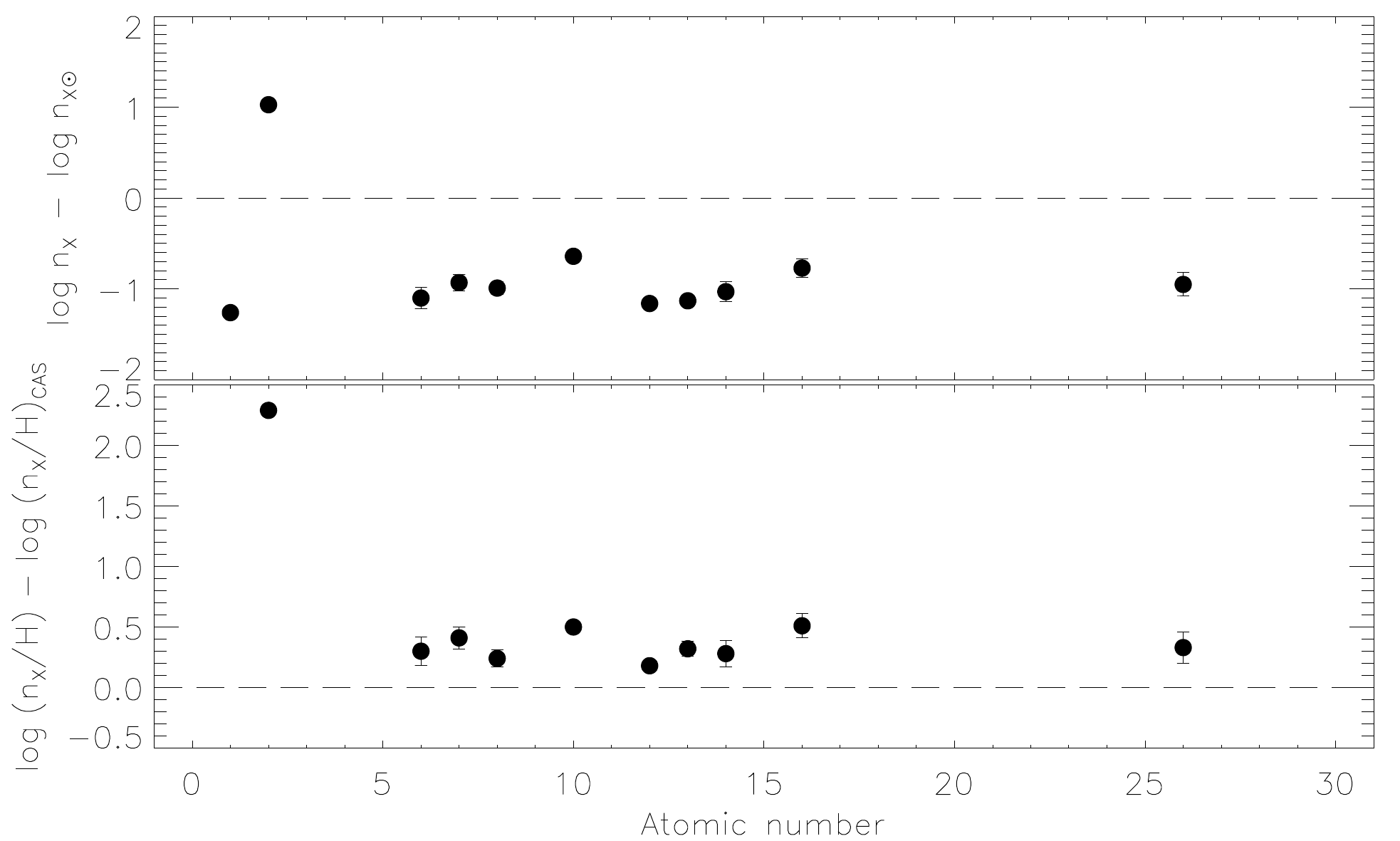}
\caption{Upper panel: Elemental abundances $\log n_X$ of HD144941 (dots)
compared to solar values \citep{Asplundetal09}. Lower panel: Elemental
abundances $\log (n_X/$H) relative to cosmic abundances \citep{NiPr12,Przybillaetal13}.
The dashed lines mark the identity.
\label{fig:abundances}}
\end{figure}

Detailed abundances from our line-by-line analysis are summarised in
Table~\ref{tab:abundances_lines_nlte}. For each spectral line, the wavelength $\lambda$ is stated, as well as the
excitation energy $\chi$ of the lower level of the transition, the adopted oscillator strength $\log gf$, an
accuracy flag for it and its source, and the non-LTE abundance value $n_X$. Average values are given in
Table~\ref{tab:parameters}, with the number of analysed lines indicated in brackets. Our metal abundances 
are about a factor four higher than in previous determinations, but they are still rather low, at
around a value of about 1/10 solar (see the upper panel of Fig.~\ref{fig:abundances}).
Carbon is normal, in contrast to the typical EHe stars where the carbon number fraction is of the 
order of 1\% (i.e. $\sim$50$\times$solar) due to admixture of $3\alpha$-process ash. Nitrogen is slightly
enriched relatively to carbon and oxygen, which may be a sign for mixing with small amounts of CN-cycled matter.
Some $\alpha$-elements like neon or sulphur are overabundant with respect to iron, which has been interpreted 
in the past as $\alpha$-enhancement as typically found in thick-disc or halo stars. 

However, a change of perspective
brings new insight. If the metal abundances are normalised to the hydrogen abundance and compared to 
abundances found for young stars in the solar neighbourhood \citep[the cosmic abundance standard,
CAS;][]{NiPr12,Przybillaetal13}, the values appear super-solar by about a factor of two (see the lower panel 
of Fig.~\ref{fig:abundances}). This is what one would expect to find for a young star from the inner Milky
Way. The canonical mechanism for producing the helium enrichment of the atmospheric layers of
He-strong stars via a weak, fractionated stellar wind in the presence of a magnetic field \citep{HuGr99}
could indeed lead to the observed pattern in
absolute numbers. The metal-line-driven wind drags the bulk plasma with it, but while Coulomb coupling
efficiently accelerates the ionised hydrogen, the neutral helium falls back to the atmosphere.
Over time, helium accumulates\footnote{This produces a negative $\mu$-gradient ($\mu$ being the mean 
molecular weight) which is unstable against turbulent motions
that might transport material of pristine composition to the surface, thus removing the chemical peculiarity. 
However, a magnetic field suppresses turbulence.} and reduces the absolute abundances of the other 
chemical constituents. The star HD144941 is special insofar as the process occurred to a much more pronounced 
extent than in any other He-strong star known.

\begin{figure*}[th]
\sidecaption
\centering
\includegraphics[width=12cm]{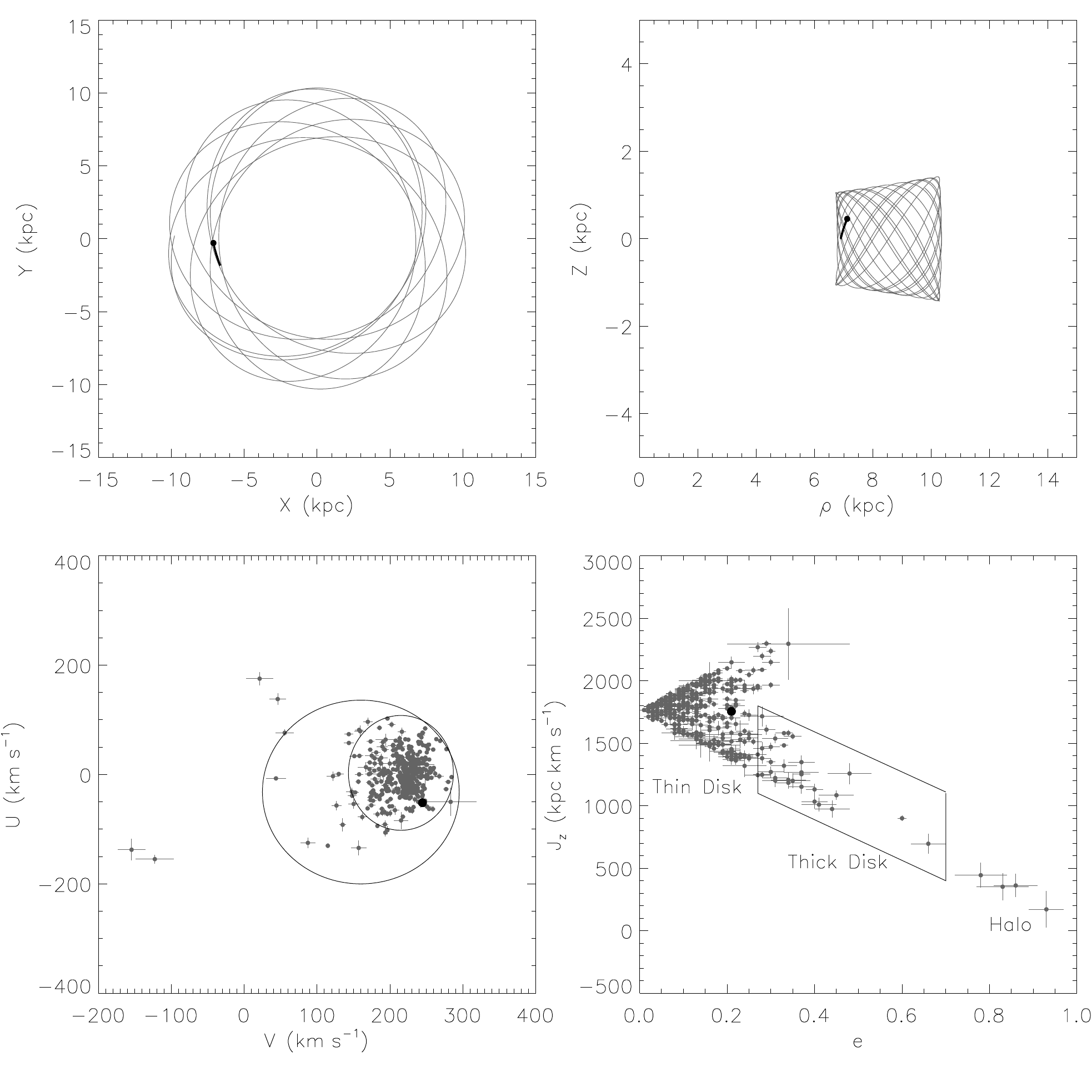}
\caption{Kinematics of HD144941 (black dots for present values) with respect to the kinematic data for
the white dwarf sample (grey dots with error bars) of \citet{Paulietal06}. Upper left panel: Orbital
motion of HD144941 in the Galactic plane. Upper right panel: Meridional orbit. Backward integration 
by about 6.2\,Myr to the Galactic equatorial plane crossing is indicated by the full black line,
integration from the present to +2\,Gyr by the grey line. Lower left panel: $V$--$U$ velocity plot.
The black ellipses render the 3$\sigma$ thin (inner) and thick disc (outer) contours as specified by
\citet{Paulietal06}. Lower right panel: $e$–-$J_z$ diagram. The solid box marks the thick-disc region as
specified by \citet{Paulietal06}, separating the thin disc and the halo region.}
\label{fig:kinematics}
\end{figure*}

\section{Kinematic analysis}
We employed the approach of \citet{Paulietal06} for the computation of the orbit and the
kinematic parameters using the code of \citet{OdBr92}. The Galactic potential of \citet{AlSa91} was
used. Parallax $\pi_\mathrm{EDR3}$ and proper motion data $\mu_\alpha$
and $\mu_\delta$ from Gaia Early Data Release 3 
\citep[EDR3][]{Gaia16,Gaia21} were utilised. The barycentric radial velocity $v_\mathrm{rad}$ was measured from the
FEROS spectrum. The Gaia-based distance $d_\mathrm{Gaia}$ was determined adopting a Gaia EDR3 parallax
zero-point offset by \citet{Renetal21}, and the spectroscopic distance following \citet{NiPr12}. The kinematic data and distances are also collected in Table~\ref{tab:parameters}.

Based on the Gaia EDR3 parallax and proper motions the geometric form of the orbit of HD144941 appears to be
thick-disc-like both in the Galactic plane and the meridional projection ($X$, $Y$, and $Z$ are
Cartesian Galactic coordinates with the Galactic centre at the origin,
and $\rho$ is the Galactocentric radius;
see the grey curve in both upper panels of Fig.~\ref{fig:kinematics}, which was calculated 2\,Gyr into the
future). However, the velocity components of the star in the radial and Galactic rotation direction ($U$, $V$) do
not match those of typical thick-disc stars \citep[according to the criteria employed by][]{Paulietal06},
and neither do the eccentricity and $Z$-component of the angular momentum $J_Z$ (lower panels of
Fig.~\ref{fig:kinematics}), which are thin-disc-like. This contradiction can be resolved if
HD144941 is a {\em \emph{runaway star}}, ejected about 6.2\,Myr ago from the Sagittarius spiral arm
in the Galactic plane (black lines), either by dynamical ejection or by a
supernova exploding in a binary system. During its remaining lifetime, HD144941 will climb in Galactic 
latitude until presumably exploding in a supernova about 1.1\,kpc above 
the Galactic plane, possibly leaving a magnetar as remnant.

\section{HD144941 in the evolutionary context}

\begin{figure}[t]
\centering
\includegraphics[width=.98\hsize]{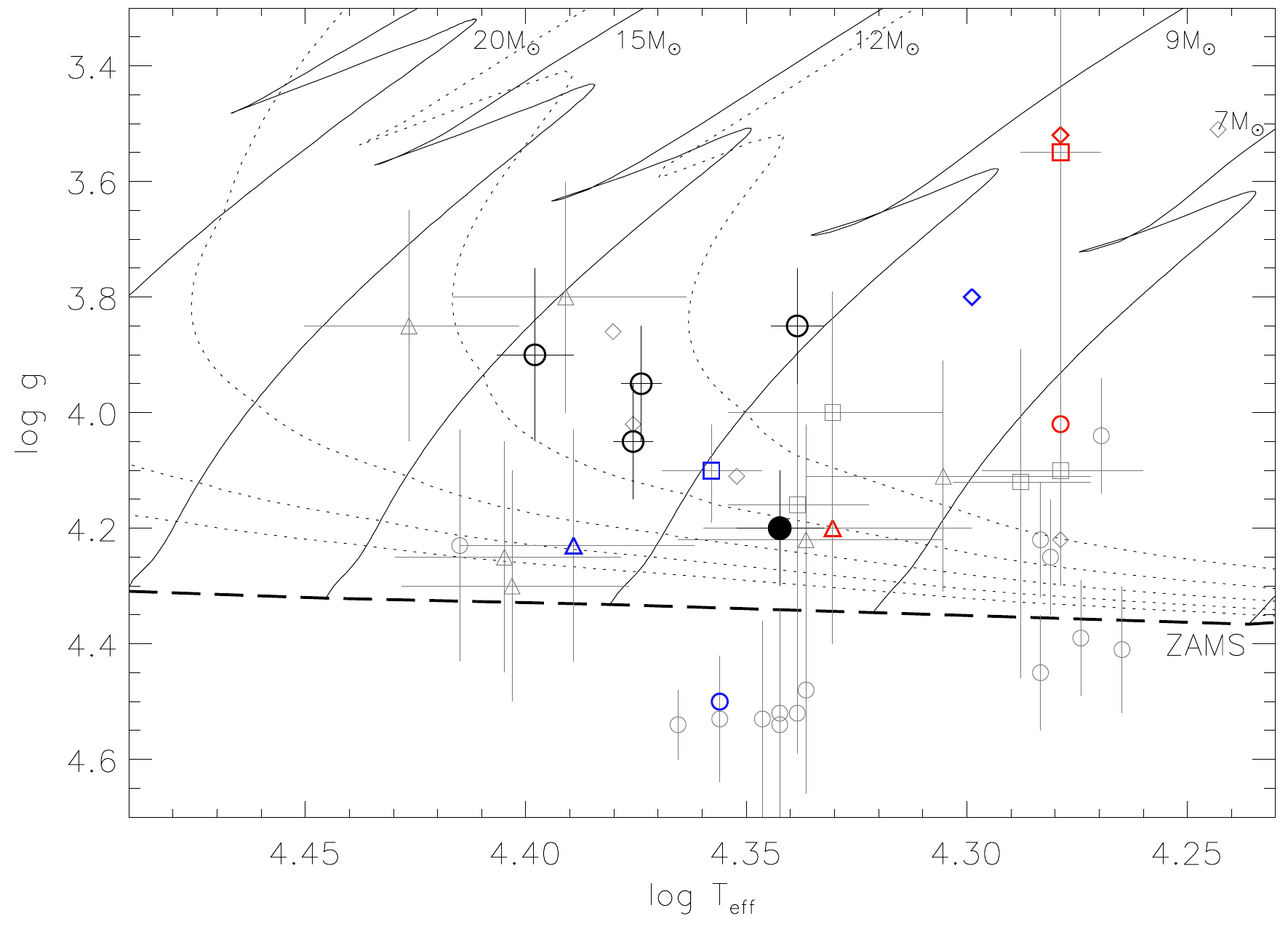}
\caption{Kiel diagram for HD144941 (black dot). Geneva evolution tracks for rotating stars are shown for 
metallicity $Z$\,=\,0.014 and an initial value of 40\% of the critical rotation rate \citep[][full 
lines]{Ekstroemetal12}. Corresponding isochrones for $\log \tau$\,(Myr)\,=\,6.5 to 7.3, with 0.2 spacing are
also  indicated (dotted lines, bottom to top), as well as the position of the zero-age main sequence (long
dashed line). Positions of other He-strong stars analysed in analogy to the present work are shown (open black 
circles). Literature values for 
He-strong stars are marked by grey symbols: from \citet[circles]{Zboriletal97}, 
\citet[diamonds]{Leoneetal97}, \citet[squares]{HuGr99} and \citet[triangles]{Cidaleetal07}. Two stars
common to all four literature analyses are highlighted, HD58260 (red symbols)
and HD64740 (blue symbols). Where available, 1$\sigma$-error bars are displayed.}
\label{fig:kieldiagram}
\end{figure}

Finally, fundamental stellar parameters such as
mass $M$, radius $R$, luminosity $L$, age $\tau,$
and ratio of age-to-main-sequence lifetime $\tau/\tau_\mathrm{MS}$ were determined
following \citet{NiPr14}. The Gaia EDR3 parallax implies beyond any doubt that HD144941 is a 
luminous massive star and not of low mass,
despite the similar $L/M$-ratio found for the object if it were an EHe star. As a consequence,
evolution tracks like those of the Geneva group \citep{Ekstroemetal12} are appropriate for constraining 
the stellar mass (and age) from a Kiel diagram to 8.1$\pm$0.3\,$M_\odot$, see Fig.~\ref{fig:kieldiagram}. 

When compared to the positions of other He-strong stars as collected from the 
literature \citep[e.g.][]{Zboriletal97,Leoneetal97,HuGr99,Cidaleetal07} (see also
\citet{Ghazaryanetal19} for a recent compilation \mbox{),} HD144941 does not appear to be distinguished 
at first glance (see Fig.~\ref{fig:kieldiagram}). However, the two stars \object{HD58260} and \object{HD64740} 
that were analysed in all four literature studies yield $T_\mathrm{eff}$-values different by many thousands of 
Kelvin, and $\log g$-values that span the main-sequence band, for the same star. Apparently, systematic 
uncertainties of the different approaches can be much larger than indicated by the
respective error bars. Moreover, about one-third of the literature values locate the stars below the zero-age 
main sequence (ZAMS), a characteristic that is unexpected for objects of chemically-normal composition except 
for their atmospheric layers.
We conclude from this that a meaningful comparison may not be feasible by considering these previous 
studies -- obviously, more tailored studies using modern models and analysis techniques are required that 
reproduce the observed spectra and other indicators simultaneously. In view of this, only four other He-strong 
stars analysed in analogy to here \citep{Przybillaetal16,Castroetal17,Gonzalezetal17,Gonzalezetal19} remain for 
a comparison, such that no firm conclusions can be drawn on the reason for HD144941's
extreme helium enrichment. We may speculate that because of the lower $T_\mathrm{eff}$
(less radiation pressure), its higher surface gravity, and much stronger magnetic field,
the He fallback is more efficient and a large chemically peculiar cap is formed around the 
magnetic pole that dominates the visible hemisphere (see Sect.~\ref{sect:bfield}).

The stellar radius can be then determined from stellar mass and surface gravity. Together with the rotation
period from the K2 light curve, the equatorial rotational velocity is determined as $\sim$14\,km\,s$^{-1}$.
Our constraint on $v \sin i$ is not very accurate because of the degeneracy with macroturbulence. As a
consequence, the inclination angle $i$ between the rotational axis and line of sight is only loosely constrained
to values between $\sim$9 and 60\degr\ (1$\sigma$-range). The luminosity can finally be determined 
from $R$ and $T_\mathrm{eff}$, which is in excellent agreement with $M_\mathrm{bol}$ determined previously. 
The age and the main-sequence lifetime  are deduced from the evolutionary tracks. All fundamental parameter 
data are incorporated in Table~\ref{tab:parameters}.

\section{Concluding remarks}
\begin{figure}
\centering
\includegraphics[width=.9\hsize]{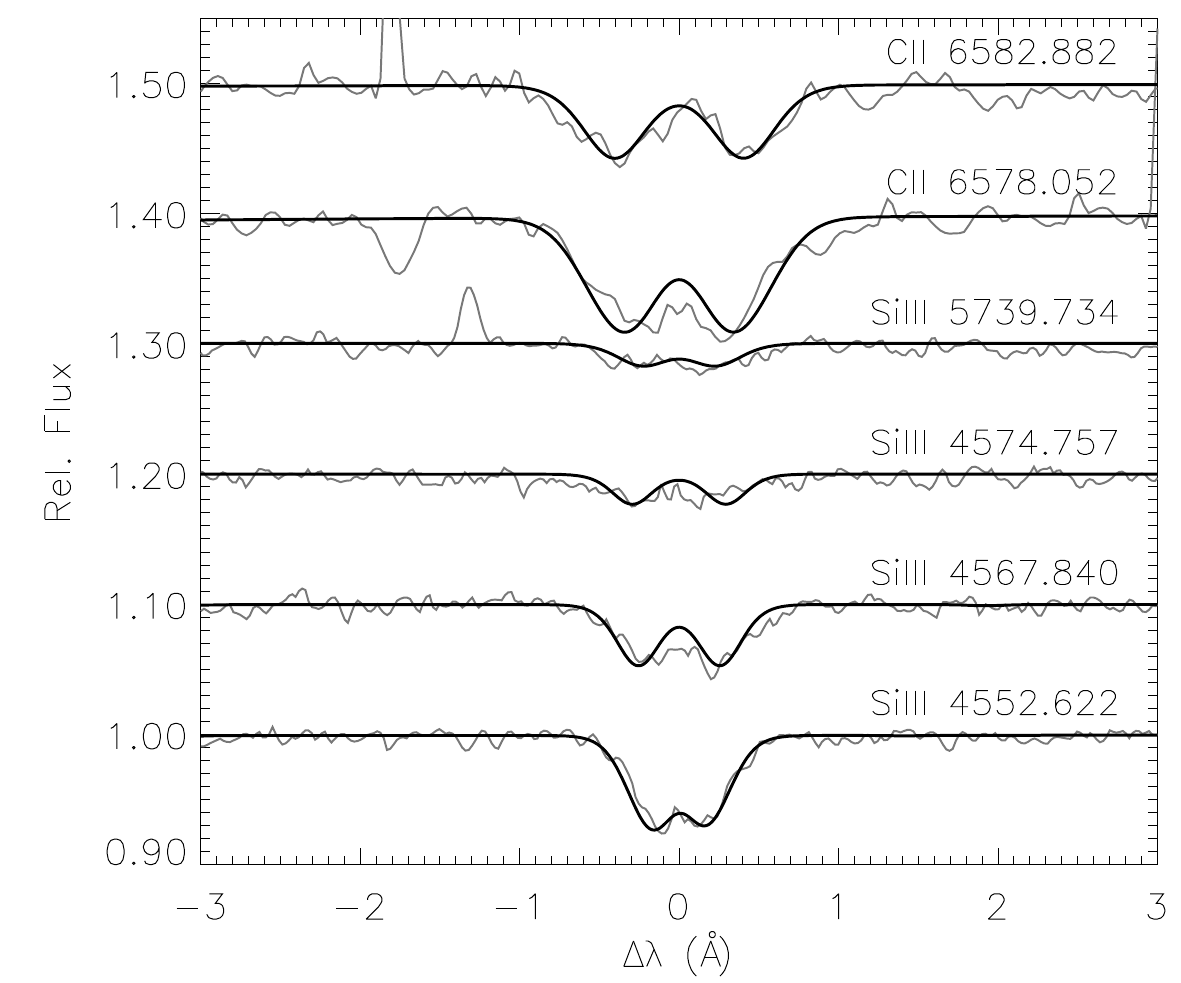}
\caption{Modelling (black line) of the Zeeman splitting assuming a longitudinal field of 15kG for several
observed spectral features (grey lines). Displayed are relative fluxes, each offset by 0.1 units, as a 
function of wavelength difference relative to the line centre. A line identification and the laboratory 
wavelengths are indicated for each feature. }
\label{fig:Zeeman}
\end{figure}

Finally, we resume the discussion of the slightly discrepant fits to some of the observed spectral
lines, the smaller model widths of the red \ion{He}{i} lines than observed, and the mismatch of the fits 
to the cores of some metal lines (while others appear to be fitted fine). 
The key to understanding lies in the \ion{C}{ii} $\lambda\lambda$6578/82\,{\AA} multiplet. 
Both lines show a pronounced 'W'-shaped profile, which is also consistently seen in other strong metal 
lines, although often barely above the noise level (see Fig.~\ref{fig:Zeeman}). 
The discrepancies in the widths of the \ion{He}{i} line cores in the red can also be largely resolved by 
assuming them to stem from Zeeman splitting.
We model the Zeeman splitting in our line-formation calculations by applying shifts and relative component
strengths as described by \citet{Sobelman92}. A good fit is obtained in Fig.~\ref{fig:Zeeman} for a field 
strength of 15\,kG by assuming a pure longitudinal field configuration;  only the $\sigma$-components are 
visible and the $\pi$-components are essentially absent from the observations of the metal lines. In addition to the
spectropolarimetry, this puts an independent constraint on the minimum magnetic field strength. 
The Zeeman splitting results in an extra line broadening $\propto$\,$\lambda^2$ (as observed 
in the red \ion{He}{i} lines), which for most lines is not 
resolved and can be accounted for by an increase in the macroturbulent velocity.
Elemental abundances are not significantly affected for most of the features because the metal lines 
are very weak and formed on the linear part of the curve of growth; that is, the splitting does not lead to
de-saturation. 
For the case where we account for the Zeeman splitting, $\zeta$ can be reduced from 25 to
15\,km\,s$^{-1}$.

A parallel work on HD144941, \citet{Shultzetal21} find very similar magnetic properties for the star from 
high-resolution spectropolarimetry and spectroscopy. Moreover, a detection of weak H$\alpha$ emission is 
reported for the first time, suggesting the existence of a centrifugal magnetosphere \citep{Petitetal13}. 
Using the rigidly rotating magnetosphere model of \citet{ToOw05}, they
propose this as strong evidence of the star having a mass around 1\,$M_{\odot}$, which 
they derive from adopting the radius determined from the Gaia EDR3 parallax and the low surface gravity 
value of \citet{PaLa17} that we ruled out here. Interestingly, \citet{Shultzetal21} note that their model 
could be reconciled with the massive star nature of HD144941 in important aspects if $v \sin i$ were lower 
than what they derived (11.2$\pm$0.7\,km\,s$^{-1}$), requiring about 7\,km\,s$^{-1}$ (c.f.~our value 
in Table~\ref{tab:parameters}, though unfortunately with a large uncertainty). They recognise their sparse 
rotational phase coverage as a potential caveat in their discussion, such that we remain optimistic that 
further observations will allow a fully consistent picture to be derived. 

We conclude that all available observational constraints investigated here imply that HD144941 is the most
extreme member of the class of He-strong stars (main-sequence B stars where atmospheric hydrogen and metals 
have been substantially depleted by a fractionated stellar wind), and that it not a member of the EHe-class of stars 
(low-mass stellar remnants with atmospheres comprising nuclear ash from CNO-cycle hydrogen burning and 
3$\alpha$ helium burning). The star HD144941 is unique with regard to helium content in the visible hemisphere, 
with almost all of the atmospheric hydrogen substituted by helium. 
Follow-up time-series spectropolarimetry at high spectral resolution would be desirable to 
investigate the magnetosphere in depth and to study the
detailed properties of the magnetic field geometry and spotted distribution of chemical elements on the
star's surface via Zeeman Doppler tomography. 

\begin{acknowledgements}
Based on observations collected at the European Southern Observatory under ESO programmes 077.D-0458(A,B), 105.205U.001 and 106.211U.001. We are grateful to K. Butler for valuable discussion and providing and maintaining {\sc Detail/Surface}, and to A. Irrgang for updates of the codes.
\end{acknowledgements}

%
%

\bibliographystyle{aa}
\bibliography{biblio.bib}

\begin{appendix}

\section{Line abundances}

\begin{table}[h!]
\caption{Line-by-line abundances from the non-LTE analysis of HD144941. Abundances $\log n_X$
(normalised to $\sum n_X$\,=\,1).
\label{tab:abundances_lines_nlte}}
\centering
\setlength{\tabcolsep}{1.5mm}
{\small
\begin{tabular}{llrrlrr}
\hline\hline
Line & $\lambda$\,({\AA}) & $\chi$(eV) &$\log gf$ & Accu. & Source &  $\log n_X$ \\
\hline
H$\delta$     & 4101.742 & 10.20 & $-$0.75243 & AAA & WF  & $-$1.30\\
H$\gamma$     & 4340.471 & 10.20 & $-$0.44666 & AAA & WF  & $-$1.28\\
H$\beta$      & 4861.333 & 10.20 & $-$0.01996 & AAA & WF  & $-$1.33\\
H$\alpha$     & 6562.819 & 10.20 &    0.71000 & AAA & WF  & $-$1.30\\[1mm]
\ion{C}{ii}   & 3918.968 & 16.33 & $-$0.533   & B   & WFD & $-$4.75\\
\ion{C}{ii}   & 3920.681 & 16.33 & $-$0.232   & B   & WFD & $-$4.62\\
\ion{C}{ii}   & 4267.001 & 18.05 &    0.562   & C+  & WFD & $-$4.83\\
\ion{C}{ii}   & 4267.261 & 18.05 &    0.717   & C+  & WFD &  \\
\ion{C}{ii}   & 4267.261 & 18.05 & $-$0.584   & C+  & WFD &  \\
\ion{C}{ii}   & 5132.947 & 20.70 & $-$0.211   & B   & WFD & $-$4.52\\
\ion{C}{ii}   & 5133.282 & 20.70 & $-$0.178   & B   & WFD &  \\
\ion{C}{ii}   & 5143.495 & 20.70 & $-$0.212   & B   & WFD & $-$4.54\\
\ion{C}{ii}   & 5145.165 & 20.71 &    0.189   & B   & WFD & $-$4.61\\
\ion{C}{ii}   & 5151.085 & 20.71 & $-$0.179   & B   & WFD & $-$4.54\\[1mm]
\ion{N}{ii}   & 3994.997 & 18.50 &    0.163   & B   & FFT & $-$5.11\\ 
\ion{N}{ii}   & 4041.310 & 23.14 &    0.748   & B   & MAR & $-$4.91\\
\ion{N}{ii}   & 4236.927 & 23.24 &    0.410   & X   & K14 & $-$5.10\\
\ion{N}{ii}   & 4237.047 & 23.24 &    0.319   & X   & K14 & \\
\ion{N}{ii}   & 4241.755 & 23.24 &    0.142   & X   & K14 & $-$5.12\\
\ion{N}{ii}   & 4241.786 & 23.24 &    0.657   & X   & K14 & \\
\ion{N}{ii}   & 4447.030 & 20.41 &    0.221   & B   & FFT & $-$4.99\\
\ion{N}{ii}   & 4601.478 & 18.47 & $-$0.452   & B+  & FFT & $-$5.00\\
\ion{N}{ii}   & 4607.153 & 18.46 & $-$0.522   & B+  & FFT & $-$5.09\\
\ion{N}{ii}   & 4621.393 & 18.47 & $-$0.538   & B+  & FFT & $-$5.08\\
\ion{N}{ii}   & 4630.539 & 18.48 &    0.080   & B+  & FFT & $-$5.13\\
\ion{N}{ii}   & 4643.086 & 18.48 & $-$0.371   & B+  & FFT & $-$5.07\\
\ion{N}{ii}   & 5001.474 & 20.65 &    0.435   & B   & FFT & $-$5.07\\
\ion{N}{ii}   & 5005.150 & 20.67 &    0.587   & B   & FFT & $-$4.91\\
\ion{N}{ii}   & 5666.629 & 18.47 & $-$0.104   & B+  & MAR & $-$4.99\\
\ion{N}{ii}   & 5676.017 & 18.46 & $-$0.356   & B+  & MAR & $-$5.02\\
\ion{N}{ii}   & 5679.558 & 18.48 &    0.221   & B+  & MAR & $-$5.18\\[1mm]
\ion{O}{i}    & 7771.944 &  9.15 &    0.354   & A   & FFT & $-$4.20\\
\ion{O}{ii}   & 3945.038 & 23.42 & $-$0.711   & B+  & FFT & $-$4.12\\
\ion{O}{ii}   & 3954.362 & 23.42 & $-$0.402   & B+  & FFT & $-$4.27\\
\ion{O}{ii}   & 4069.623 & 25.63 &    0.144   & B+  & FFT & $-$4.33\\
\ion{O}{ii}   & 4069.882 & 25.64 &    0.352   & B+  & FFT & \\
\ion{O}{ii}   & 4072.153 & 25.65 &    0.528   & B+  & FFT & $-$4.33\\
\ion{O}{ii}   & 4075.862 & 25.67 &    0.693   & B+  & FFT & $-$4.13\\
\ion{O}{ii}   & 4085.112 & 25.65 & $-$0.191   & B+  & FFT & $-$4.24\\
\ion{O}{ii}   & 4153.298 & 25.84 &    0.070   & B+  & FFT & $-$4.15\\
\ion{O}{ii}   & 4317.139 & 22.97 & $-$0.368   & B+  & FFT & $-$4.25\\
\ion{O}{ii}   & 4319.630 & 22.98 & $-$0.372   & B+  & FFT & $-$4.32\\
\ion{O}{ii}   & 4345.560 & 22.98 & $-$0.342   & B+  & FFT & $-$4.33\\
\ion{O}{ii}   & 4349.426 & 23.00 &    0.073   & B+  & FFT & $-$4.38\\
\ion{O}{ii}   & 4351.260 & 25.66 &    0.202   & B+  & FFT & $-$4.32\\
\ion{O}{ii}   & 4351.457 & 25.66 & $-$1.013   & B   & FFT & \\
\ion{O}{ii}   & 4366.892 & 23.00 & $-$0.333   & B+  & FFT & $-$4.31\\
\ion{O}{ii}   & 4414.905 & 23.44 &    0.207   & B   & FFT & $-$4.26\\
\ion{O}{ii}   & 4416.974 & 23.42 & $-$0.043   & B   & FFT & $-$4.22\\
\ion{O}{ii}   & 4590.974 & 25.66 &    0.331   & B+  & FFT & $-$4.28\\
\ion{O}{ii}   & 4595.957 & 25.66 & $-$1.022   & B   & FFT & $-$4.27\\
\ion{O}{ii}   & 4596.175 & 25.66 &    0.180   & B+  & FFT & \\
\ion{O}{ii}   & 4638.856 & 22.97 & $-$0.324   & B+  & FFT & $-$4.21\\
\ion{O}{ii}   & 4641.810 & 22.98 &    0.066   & B+  & FFT & $-$4.30\\
\ion{O}{ii}   & 4649.135 & 23.00 &    0.324   & B+  & FFT & $-$4.31\\
\ion{O}{ii}   & 4650.839 & 22.97 & $-$0.349   & B+  & FFT & $-$4.25\\[1mm]
\ion{Ne}{i}   & 6143.063 & 16.62 & $-$0.070   & B+  & FFT & $-$4.70\\
\ion{Ne}{i}   & 6402.248 & 16.62 &    0.365   & B+  & FFT & $-$4.65\\
\hline
\end{tabular}
}
\end{table}

\begin{table}[ht!]
\setcounter{table}{0}
\caption{(cont.)}
\centering
\setlength{\tabcolsep}{1.5mm}
{\small
\begin{tabular}{llrrlrl}
\hline\hline
Line & $\lambda$\,({\AA}) & $\chi$(eV) &$\log gf$ & Accu. & Source &  $\log n_X$ \\
\hline
\ion{Mg}{ii}  & 4481.126 &  8.86 &    0.730   & B   & FW  & $-$5.52\\
\ion{Mg}{ii}  & 4481.150 &  8.86 & $-$0.570   & B   & FW  &  \\
\ion{Mg}{ii}  & 4481.325 &  8.86 &    0.575   & B   & FW  &  \\[1mm]
\ion{Al}{iii} & 4512.565 & 17.81 &    0.408   & A+  & FFTI& $-$6.57\\ 
\ion{Al}{iii} & 4528.945 & 17.82 & $-$0.291   & A+  & FFTI& $-$6.67\\
\ion{Al}{iii} & 4529.189 & 17.82 &    0.663   & A+  & FFTI& $-$6.71\\
\ion{Al}{iii} & 5696.604 & 15.64 &    0.232   & A+  & FFTI& $-$6.72\\[1mm]
\ion{Si}{ii}  & 3856.018 &  6.86 & $-$0.442   & C+  & FFTI& $-$5.59\\
\ion{Si}{ii}  & 3862.595 &  6.86 & $-$0.757   & C+  & KP  & $-$5.32\\
\ion{Si}{ii}  & 4128.054 &  9.84 &    0.359   & B   & KP  & $-$5.46\\
\ion{Si}{ii}  & 4130.872 &  9.84 & $-$0.783   & D+  & KP  & $-$5.57\\
\ion{Si}{ii}  & 4130.894 &  9.84 &    0.552   & B   & KP  & \\
\ion{Si}{ii}  & 5041.024 & 10.07 &    0.17    & D+  & WSM & $-$5.53\\
\ion{Si}{ii}  & 5055.984 & 10.07 &    0.42    & D+  & WSM & $-$5.62\\
\ion{Si}{ii}  & 5056.317 & 10.07 & $-$0.53    & E   & WSM &  \\
\ion{Si}{ii}  & 6347.109 &  8.12 &    0.177   & B+  & FFTI& $-$5.34\\
\ion{Si}{ii}  & 6371.371 &  8.12 & $-$0.126   & C+  & FFTI& $-$5.43\\
\ion{Si}{iii} & 4552.622 & 19.02 &    0.292   & C+  & FFTI& $-$5.57\\
\ion{Si}{iii} & 4567.840 & 19.02 &    0.068   & B+  & FFTI& $-$5.57\\
\ion{Si}{iii} & 4574.757 & 19.02 & $-$0.409   & B   & FFTI& $-$5.64\\
\ion{Si}{iii} & 4813.333 & 25.98 &    0.708   & B+  & KP   & $-$5.42\\
\ion{Si}{iii} & 4819.631 & 25.98 & $-$1.841   & X   & K13  & $-$5.38\\
\ion{Si}{iii} & 4819.712 & 25.98 &    0.823   & B+  & KP  & \\
\ion{Si}{iii} & 4819.814 & 25.98 & $-$0.353   & B   & KP  & \\
\ion{Si}{iii} & 4828.950 & 25.99 &    0.937   & B+  & KP  & $-$5.38\\
\ion{Si}{iii} & 4829.029 & 25.99 & $-$3.169   & X   & K13 & \\
\ion{Si}{iii} & 4829.111 & 25.99 & $-$0.353   & B   & KP  & \\
\ion{Si}{iii} & 4829.214 & 25.99 & $-$2.170   & D+  & KP  & \\
\ion{Si}{iii} & 5739.734 & 19.72 & $-$0.078   & B   & FFTI& $-$5.40\\[1mm]
\ion{S}{ii}   & 4162.665 & 15.94 &    0.78    & D   & WSM & $-$5.67\\
\ion{S}{ii}   & 5432.797 & 13.62 &    0.205   & C+  & FFTI& $-$5.54\\
\ion{S}{ii}   & 5453.855 & 13.67 &    0.442   & C+  & FFTI& $-$5.52\\
\ion{S}{iii}  & 4253.499 & 18.24 &    0.107   & B+   &FFTI& $-$5.74\\
\ion{S}{iii}  & 4284.904 & 18.19 & $-$0.233   & B+   &FFTI& $-$5.54\\[1mm]
\ion{Fe}{iii} & 4164.731 & 20.63 &    0.923   & X   & KB  & $-$5.54\\
\ion{Fe}{iii} & 4164.916 & 24.65 &    1.011   & X   & KB  & \\
\ion{Fe}{iii} & 4372.039 & 22.91 &    0.585   & X   & KB  & $-$5.36\\
\ion{Fe}{iii} & 4372.096 & 22.91 &    0.029   & X   & KB  & \\
\ion{Fe}{iii} & 4372.134 & 22.91 &    0.727   & X   & KB  & \\
\ion{Fe}{iii} & 4372.306 & 22.91 &    0.865   & X   & KB  & \\
\ion{Fe}{iii} & 4372.306 & 22.91 &    0.193   & X   & KB  & \\
\ion{Fe}{iii} & 4372.497 & 22.91 &    0.200   & X   & KB  & \\
\ion{Fe}{iii} & 4372.536 & 22.91 &    0.993   & X   & KB  & \\
\ion{Fe}{iii} & 4372.784 & 22.91 &    0.040   & X   & KB  & \\
\ion{Fe}{iii} & 4372.823 & 22.91 &    1.112   & X   & KB  & \\
\ion{Fe}{iii} & 4419.596 &  8.24 & $-$2.218   & X   & KB  & $-$5.39\\
\ion{Fe}{iii} & 5127.387 &  8.66 & $-$2.218   & X   & KB  & $-$5.22\\
\ion{Fe}{iii} & 5127.631 &  8.66 & $-$2.564   & X   & KB  & \\
\ion{Fe}{iii} & 5156.111 &  8.64 & $-$2.018   & X   & KB  & $-$5.53\\
\hline
\end{tabular}
\tablefoot{Accuracy indicators -- uncertainties within the following:
AAA: 0.3\%;
AA: 1\%;
A: 3\%;
B: 10\%;
C: 25\%;
D: 50\%;
E: >50\%;
X: unknown. 
Sources of $gf$-values:
FFT:    \citet{FFT04};
FFTI:   \citet{FFTI06};
FW:     \citet{FuWi98};
K13, K14: {\tt http://kurucz.harvard.edu/atoms.html};
KB:     \citet{KuBe95};
KP:     \citet{KePo08};
MAR:    \citet{Maretal00};
WF:     \citet{WiFu09};
WFD:    \citet{WFD96};
WSM:    \citet{Wieseetal69}.
}}
\end{table}

\end{appendix}
\end{document}